\documentstyle[12pt]{article}
\renewcommand{\baselinestretch}{1.1}
\begin{document}
\centerline{\Huge Two body non-leptonic $\Lambda_b$ decays}
\centerline{\Huge in quark model with factorization ansatz }
\vspace{1cm}
\centerline{\bf Fayyazuddin}
\centerline{Department of Physics}
\centerline{Quaid-i-Azam University}
\centerline{Islamabad, Pakistan.}
\vspace{1cm}
\centerline{\bf Riazuddin}
\centerline{Department of Physics}
\centerline{King Fahd University of Petroleum and Minerals}
\centerline{Dhahran 31261, Saudi Arabia.}
\vspace{1cm}

\begin{abstract}
The two body non-leptonic $\Lambda_b$ decays are
analyzed in factorization approximation, using  quark model,
$\xi = 1 / N_c$ as a free parameter.
It is shown that the experimental branching
ratio for $\Lambda_b \longrightarrow \Lambda {J/\psi}$ restricts
$\xi$ and this ratio can be understood for a
value of $\xi$ which lies in the range $ 0 \leq \xi \leq 0.5 $
suggested by two body B meson decays. The branching ratios for
$\Lambda_b \longrightarrow \Lambda_{c} D^*_s(D_s) $ are predicted
to be larger than the previous estimates. Finally it  is
pointed that CKM-Wolfenstein parameter $\rho^2 + \eta^2$,
where $\eta$ is CP phase, can be
determined from the ratio of widths of
$\Lambda_b \longrightarrow \Lambda \bar{D}$ and
$\Lambda_b \longrightarrow \Lambda {J/\psi}$ or that of
$\Lambda_b \longrightarrow \, p \, D_s$ and
$\Lambda_b \longrightarrow \Lambda_c \, D_s$
independent of the parameter $\xi$.
\end{abstract}

\renewcommand{\baselinestretch}{1.5}
\newpage
\section{INTRODUCTION}
Two body non-leptonic decays of bottom baryons provide
useful information for QCD effects in weak decays
and indirect CP asymmetries which involve
CKM-Wolfenstein parameters $\rho$ and $\eta$. The standard
frame work to study non-leptonic decays of
bottom baryons is provided by an effective
Hamiltonian approach, which allows
a separation between short- and long-distance
contributions in these decays. The latter involves
the matrix elements $ < M B'| O_i | B >$ at a
typical hadronic scale, where $O_i$ is an operator
in the effective Hamitonian. These matrix elements
cannot be calculated at present from first principles.
Thus one has to resort to some approximate schemes.
Such schemes are often complicated by competing mechanisms,
such as factorization, baryon pole terms
and W-exchange terms, each of which has uncertainties of its own.
The purpose of this paper is to study a class of two body bottom
baryon non-leptonic decays in the framework of factorization scheme,
where neglecting final state interactions, hadronic matrix elements
are
factorized into a product of two matrix elements of the form
$<B'|J_\mu|B>$ and $<0|J'_\mu |M>$ for which more
information may be available.

Following the phenomenological sucess of factorization in the
heavy to heavy non-leptonic B-meson decays \cite{R1},
this frame work has been extended to the domain of heavy to light
transitions \cite{R2}.
The factorization anstaz here introduces one free
parameter, called $\xi = 1/N$ ( $N$ being number of colors), which
is introduced to compensate for the neglect of color octet-octet
contribution
in evaluating the hadronic matrix elements in the heavy to light
sectors.
The range $0\leq \xi \leq 0.5 $ has been
found \cite{R2} to be consistent with data on a number of
measured B meson decays. We apply the factorization to decays
$\Lambda_b \longrightarrow \Lambda {J/\psi}$,
$\Lambda_b \longrightarrow \Lambda_c  D_s({D^*_s})$,
$\Lambda_b \longrightarrow \Lambda \bar{D}$ and
$\Lambda_b \longrightarrow \, p \, D_s$.
In addition, we use  quark model to fix current
coupling constants which appear in the matrix elements
$<B'|J_\mu |B>$.
We show that the measured branching ratio for
$\Lambda_b \longrightarrow \Lambda {J/\psi}$ can be accounted
for in this approach with the parameter
$\xi$ in the above mentioned range. Our estimates for branching
ratios for
$\Lambda_b \longrightarrow \Lambda_c D_s (D^*_s)$
are larger   than their
previous estimates \cite{R3,R4}. The decays
$\Lambda_b \longrightarrow \Lambda \bar{D}$
and $\Lambda_b \longrightarrow \, p \, D_s$  can give information
on the CKM-Wolfenstein parameter $(\rho^2 +\eta^2 )$ \cite{R5}
or $|V_{ub}/V_{cb}|$
independent of $\xi$.

We write the effective Hamiltonian\cite{R6}:
\begin{eqnarray}
\label{eq:01} H_{eff} (\Delta B= 1) &=& \frac{G_F}{\sqrt{2}}
\left[ \sum_{q=u,c} V_{cb} V^*_{qs} \left( C_1O_1^c + C_2 O_2^c
\right)
\right. \nonumber \\
&&+ \left. \sum_{q=u,c} V_{ub} V^*_{qs}\left( C_1O_1^u + C_2 O_2^u
\right) \right]
\end{eqnarray}
where $C_i$ are Wilson coefficients evaluated at the renormalization
scale $\mu$;
the current-current operators $O_{1,2}$ are
\begin{eqnarray}
\label{eq:02}
O^c_1 &=& \left( \bar{c}^\alpha b_\alpha \right)_{V - A}
\left( \bar{s}^\beta q_\beta \right)_{V - A} \nonumber \\
O^c_2 &=& \left( \bar{c}^\alpha b_\beta \right)_{V - A}
\left( \bar{s}^\beta q_\alpha \right)_{V - A}
\end{eqnarray}
and $O^u_i$ are obtained through replacing $c$ by $u$.
Here $\alpha$ and $\beta$ are $SU(3)$ color indices while
$(\bar{c}^\alpha b_\beta)_{V - A} = \bar{c}^\alpha \gamma_\mu
(1+\gamma_5) b_\beta$
etc. The related Wilson coefficients at $\mu = 2.5 GeV$
in next-to-leading logarithmic (NLL) precision are \cite{R2}
\begin{eqnarray}
\label{eq:03}
C_1 &=& 1.117 \nonumber \\
C_2 &=& - 0.257
\end{eqnarray}
These are not very different from those at $\mu = 5 GeV $ in the
leading
logarithmic approximation (LLA) \cite{R7}: $C_1 (m_b) = 1.11$ and
$C_2 (m_b) = - 0.26$.

In the factorization scheme we encounter matrix
elements of the form
\begin{eqnarray}
\label{eq:04}
<B(p')|J_\mu |B_b(p)>  &=& \bar{u} (p') \Gamma_\mu u (p) \nonumber \\
&=& \bar{u} (p') i \left[ \left( g_V (s)
- g_A(s) \gamma_5 \right) \gamma_\mu \right. \nonumber \\
&& +  \left( f_V(s) + h_A (s) \gamma_5 \right) \sigma_{\mu \nu} q_\nu
\nonumber \\ && + \left. i \left( h_V(s)  -   f_A (s) \gamma_5
\right)q_\mu \right] u(p)
\end{eqnarray}
where $B_b$ is a baryon, which contains $b$ quark while $B$ is any
baryon not containing
it. Here $ s = - q^2 = - (p-p')^2$. In the heavy quark spin symmetry
limit \cite{R8},
the vector and axial vector form factors are related [when
$B_b$ belongs to the triplet representation of flavor SU(3)] as
follows:
\begin{equation}
\label{eq:05}
g_V(s)=g_A(s)=f_1
\end{equation}
\begin{equation}
\label{eq:06}
f_V = h_V =  h_A = - f_A = \frac{1}{m_{B_b}} f_2
\end{equation}

For a decay of
the type $B_b (p) \longrightarrow B(p') + X (p_X)$,
the matrix elements are of the form
\begin{equation}
\label{eq:07}
T=i \frac{G'}{\sqrt{2}} <0|J'_\mu|X(p_X)>\bar{u}(p') \Gamma_\mu
 u(p)  \frac{1}{(2\pi)^3}
\sqrt{\frac{mm'}{p_op'_o}}
\end{equation}
In the rest frame of $B_b$, the decay rate of $B_b$ and its
polarization are given by

\begin{equation}
\label{eq:08}
\Gamma = \frac{G'^2}{2} \frac{1}{4\pi m^2} \int ds p'(s) \left\{
\rho (s) \Gamma^\rho (s) + \sigma (s) \Gamma^\sigma (s) \right\}
\end{equation}
where $q = p_X = p - p', \,\, s = - q^2$ and
\begin{eqnarray}
\Gamma^\rho (s) &=& \left\{ Q (s) \left( g^2_V +g^2_A \right)
- 3 m m' s \left( g^2_V  - g^2_A   \right) \right. \nonumber \\
&&+ 3 s \left[ (m + m') \left( (m - m')^2 -s \right) g_V f_V \right.
\nonumber \\ && - \left. (m - m') \left( (m + m')^2 -s \right) g_A
f_A \right] \nonumber \\
&&+ s \left[ Q''(s) \left( f^2_V + h^2_A \right) - 3 m m' s
    \left( f^2_V - h^2_A \right) \right] \nonumber \\
&&- 2 m p'(s) {\bf n} \cdot {\bf s} \left[
    \left( (m^2 - m'^2) - 2s \right) g_A g_V \right. \nonumber \\
&&+ s \left( (m - 3 m') g_V h_A - (m + 3 m') g_A f_V \right)
\nonumber \\
&&+ s f_V h_A \left. \left. \left( s - m^2 - 5 m'^2 \right) \right]
\right\}
\end{eqnarray}
\begin{eqnarray}
\label{n10}
\Gamma^\sigma (s) &=& \left\{ Q' (s) \left( g^2_V +g^2_A  \right)
  -s \left[ (m - m') \left( (m + m')^2 -s \right) g_V f_V \right.
\right. \nonumber \\
&&+ \left.  (m + m') \left( (m - m')^2 -s \right) g_A f_A \right]
\nonumber \\
&&+ \frac{1}{2} s^2  \left[ \left( (m + m')^2 -s \right) h_V^2
    + \left( (m - m')^2 -s \right) f_A^2 \right] \nonumber \\
&& - 2 m p'(s) {\bf n} \cdot {\bf s} \left[ (m^2 - m'^2) g_V g_A
\right. \nonumber \\
&& \left. \left. - s \left(  (m + m') g_A h_V + (m - m') g_V g_A
\right)
 + s^2 h_V f_A \right] \right\}
\end{eqnarray}
Here it is understood that form factors are functions of $s$ and
$\rho = \rho_V , \,\, \rho_A \mbox{ or } 0$ according as $X$ is
$1^-$, $1^+$ or $O^-$, while correspondingly $\sigma = 0$,
$\sigma_A$ or $\sigma_{p\cdot s}$ and
\begin{eqnarray}
\label{eq:13}
p'(s) &=& \frac{1}{2m} { \left\{ \left[ \left( m^2 + m'^2 \right) -s
    \right]^2 - 4m^2 m'^2 \right\}^{1/2} } \\
\label{eq:11}
Q(s) &=& \frac{1}{2} \left[ \left( m^2 - m'^2 \right)^2 +
    s \left( m^2 + m'^2 \right)  - 2 s^2 \right] \\
\label{eq:12}
Q'(s) &=& \frac{1}{2} \left[ \left( m^2 - m'^2 \right)^2
    - s \left( m^2 + m'^2 \right) \right] \\
\label{n14}
Q''(s) &=& \frac{1}{2} \left[ 2 \left( m^2 - m'^2 \right)^2
    - s \left( m^2 + m'^2 \right) - s^2 \right]
\end{eqnarray}

The form factors defined in Eq.(4) are calculated in quark model at
$s =-q^2 = m_X^2$ where $X$ is a vector or pseudoscalar  particle  in
the
decay $ B \longrightarrow  B' X $, thereby taking into account recoil
correction. This is in contrast to the use of the non-relativistic
quark model
for the evaluation of form factors at zero recoil $q=0$ [9].
This latter approach also necessitates the extrapolation of form
factors
from maximum $q^2 [  - q^2_m = t_m = (m_B - m_{B'} )^2 ]$ to the
desired
$s = - q^2 = m_X^2$. We may point out that since $|{\bf q}| \approx
1.75 GeV$ in $\Lambda_b \longrightarrow \Lambda + J/ \psi $, for
example,
the no recoil approximation does not seem to be justified; in fact
$|{\bf q}|
\gg m_s$ in $\Lambda$, making the $s$ quark in $\Lambda $
relativistic.
In our approach no recoil approximation, nor any extrapolation of
form factors
at the physical point are needed. Our quark model results do satisfy
the constraints imposed by heavy quark spin symmetry.

The plan of the paper is as follows: Sec.11 summarizes the
calculation of the
baryonic form factors within the frame work of quark model at the
desired value
of $s = - q^2$, rather than at the zero recoil point, relegating the
details in the appendix.
In section III we apply the results to some specific nonleptonic
decay modes of $\Lambda_b$. Section IV summarizes our conclusions.

\section{BARYONIC FORM FACTORS IN QUARK MODEL}

In order to calculate the form factors we first reduce the matrix
elements
in Eq.(4) from four component Dirac spinors to Pauli spinors without
making any approximation and do the same for the quark level current
\begin{equation}
j_\mu = i \bar{q} \gamma_\mu ( 1 + \gamma_5) b
\end{equation}
We treat the b quark in $B_b$ extremely non relativistically
$( {\bf p}_b / m_b \approx 0)$
and set $ {\bf p}_b - {\bf p}_q = {\bf q} = - {\bf p}', \,\,\,\,
E_q
= \sqrt{{\bf q}^2 + m_q^2}$, $E_b = m_b = m_3.$ Then as shown in the
appendix:
\begin{equation}
h_A = - f_A, \,\,\,\,\, f_V = h_V
\end{equation}
\begin{eqnarray}
g_V(s) &=& \xi_V I a(E',E'_3) \nonumber \\
f_V(s) &=& \frac{1}{m} \xi_V I b(E', E'_3) \nonumber \\
g_A(s) &=& \xi_A I a(E', E'_3) \nonumber \\
f_A(s) &=& - \frac{1}{m} \xi_A I b(E',E'_3)
\end{eqnarray}
where
\begin{eqnarray}
a(E',E'_3) &=& \frac{1}{2}\sqrt{\frac{E'}{E'_3}}
\frac{(E' + m')\left( 1 - \frac{m'}{m}\right) +
(E'_3 + m'_3)\left( 1 + \frac{m'}{m}\right)}{\sqrt{(E' + m')
(E'_3 + m'_3)}}
\nonumber \\
b (E',E'_3) &=& \frac{1}{2}\sqrt{\frac{E'}{E'_3}}
\frac{(E' + m') - (E'_3 + m'_3)}{\sqrt{(E' + m') (E'_3 + m'_3)}}
\end{eqnarray}
and $E' = p'_o, \,\,\, E'_3 = E_q = \sqrt{{\bf p}'^2 + m'^2_3}$, and
$m'_3 = m_q.$
Note the explicit appearance of $1/m$ corrections in the
above formulae. Here
$\xi_V$ and $\xi_A$ are respectively the spin-unitary spin part of
the matrix elements of the current operator (13);
for example, for $B_b$ belonging to the triplet representation of
$SU(3)$, $\xi_V = \xi_A$
and $I$ is overlap integral
\begin{equation}
I = N_f N_i \int \psi^*_f \left( {\bf p}_{12},
{\bf k} - \frac{m_1 +m_2}{\tilde{m}'} {\bf p}' \right)
\psi_i ({\bf p}_{12}, {\bf k}) d^3 p_{12} d^3 k
\end{equation}
The recoil correction is represented by momentum mismatch
$\frac{m_1 +m_2}{\tilde{m}'} {\bf p}'$, which arises since the rest
frame
of $B_b$ is not that of $B_q$ baryon.
Here $\tilde{m}' = m_1 +m_2 +m_3'$ where $m_1$ and $m_2$ are masses
of the
spectator quarks
and $m'_3$ is that of $q$ quark resulting from the decay of $b$. Note
that the form factors in Eq.(15) are determined at the desired
value of $s = - q^2$.

As already noted for $B_b$ belonging to the triplet representation
$\xi_V = \xi_A$
and then the relations (14) and (15) are consistent
with those given in Eqs.(5) and (6) obtained in the heavy
quark spin symmetry limit.
To proceed further we use harmonic oscillator or Gaussain
wave functions in Eq.(20)to obtain
\begin{equation}
I = \left( \frac{2 \beta \beta'}{\beta^2 + \beta'^2} \right)^3
\exp \left[ - \frac{3}{4} \frac{(m_1 +m_2)^2}{\tilde{m}'^2}
\frac{p'^2}{2 (\beta^2 + \beta'^2)} \right]
\end{equation}
We take $\beta$ or $\beta'$ as [10]
\begin{equation}
\beta^2 = \sqrt{\mu_Q \kappa }
\end{equation}
where $\mu_Q = \frac{M_N M_H}{M_N + M_H}$ is the reduced mass of
the bound system, $M_N$ being the nucleon mass and $M_H$ that of
$B$, $D$, $K^*$ or $\rho$ meson for $\Lambda_b$, $\Lambda_c$,
$\Lambda$ and $p$ respectively. $\kappa$ is the spring constant
and its value is taken to be $(440 \, MeV)^3$ [11].

We summarize in Table 1, the form
factors $g_V(s) = g_A(s) = f_1$, $f_V(s) = h_V(s) = h_A(s) =-f_A(s)
=f_2/m$
for the transitions
$ \Lambda_b \longrightarrow pD_s$,
$ \Lambda_b \longrightarrow \Lambda D$,
$ \Lambda_b \longrightarrow \Lambda J/\psi$,
$ \Lambda_b \longrightarrow \Lambda_c D^*_s (D_s)$,
for $s = m^2_{D_s}$, $m^2_D$ $m^2_{J/\psi}$,
and $m^2_{D^*_s}$ $(m^2_{D_s})$, $m_3' = m_u$, $m_s$ and $m_c$
respectively.
For the numerical work we have taken the relevant masses ( in GeV) as
$m =m_{\Lambda_b} = 5.641$, $m_\Lambda = 1.1157$, $m_{\Lambda_c} =
2.285$,
  $m_p =0.938$, $m_{J/\psi} = 3.097$, $m_{D^*_s}=2.112$,
  $m_D =1.864$, $m_{D_s}=1.968$, $m_s = 0.510$, $m_c =1.6$ and
  $m_u=0.340$.
  
  \begin{table}
  \caption{Quark model predictions for baryonic form factors for
  $\Lambda_b $ transitions. $\beta=0.51\, GeV, \beta'=0.44\, GeV$ for
  $p$ and $\Lambda$ and $=0.48\, GeV$  for $\Lambda_c$. $f_1=g_V=g_A$,
  $f_2/f_1=(f_V/g_V)m=(h_V/g_V)m=(h_A/g_A)m =-(f_A/g_A)m$. Note that
  only the last column depends on the overlap integral $I$.}
  
  \begin{tabular}{ccccccccc} \hline \hline
  Transition&$|p'|$&$\xi_A=\xi_V$&$f_1(s)/I$&$f_2/f_1$&${\cal
  F}_1$&${\cal F}_2$&$I$&$f_1(s)$ \\ \hline
  $p D_s$&2.376&$1/\sqrt{2}$&0.720&0.123&$\approx 1$&$\approx
  1$&0.119&0.086 \\ \hline
  $\Lambda D$&2.374&$-1/\sqrt{3}$&-0.558&0.129&$\approx 1$&$\approx
  1$&0.215&-0.120 \\ \hline
  $\Lambda J/\psi$&1.756&$-1/\sqrt{3}$&-0.604&0.158&0.943&0.826&0.426&-
  0.257 \\ \hline
  $\Lambda_c D_s$&1.766&1&1.052&0.134&0.978&0.983&0.791&0.829 \\ \hline
  $\Lambda_c D_s^*$&1.850&1&1.048&0.137&0.949&0.908&0.810&0.852 \\
  \hline \hline
  \end{tabular}
  \end{table}
  
  \section{APPLICATIONS}
  
  We consider those decays of $\Lambda_b $ for which baryon poles
  either
  do not contribute or their contribution is highly suppressed due to
  Okubo-Zweig-Iizuka (OZI) rule  and that it scales as inverse of
  $m_{\Lambda_b}$.
  
  For decays of type $\Lambda_b(p) \longrightarrow B_q(p') V(q)$,
  where $V$ is a vector meson,
  \begin{equation}
  \rho_A (s) = 0 = \sigma_A (s)
  \end{equation}
  \begin{equation}
  \rho_V(s) = F^2_V \delta (s-m^2_V)
  \end{equation}
  where
  \begin{equation}
  <0|J'_\mu|V> = F_V \epsilon_\mu
  \end{equation}
  Then Eqs.(9) and (10), on using the relations (5), give the decay rate
  \begin{equation}
  \Gamma = \frac{G'^2}{2} F_V^2  \frac{|{\bf p'}|}{4\pi m^2}
  Q(m_V^2) [2f^2_1(m_V^2)]{\cal F}^V_1(m_V^2)
  \end{equation}
  while the asymmetry
  \begin{equation}
  \alpha = \frac{-2m|{\bf p'}| \left[ (m^2 - m'^2) - 2m_V^2
  \right]}{2Q(m_V^2)}
  \frac{{\cal F}^V_2(m_V^2)}{{\cal F}^V_1(m_V^2)}
  \end{equation}
  where
  \begin{equation}
  {\cal  F}^V_1(m_V^2) = \left\{ 1 - 3 \frac{m'}{m} \frac{m_V^2 (m^2-
  m'^2 +m_V^2)}{Q(m_V^2)}
  \frac{f_2}{f_1} + \frac{m_V^2}{m^2} \frac{Q''(m_V^2)}{Q(m_V^2)}
  \frac{f^2_2}{f_1^2} \right\} \nonumber \\
  \end{equation}
  \begin{equation}
  {\cal F}^V_2(m_V^2) = \left\{ 1-6
  \frac{m'}{m}\frac{m_V^2}{m^2-m'^2-2m_V^2}\frac{f_2}{f_1}
  -\frac{m_V^2}{m^2}\frac{m^2 +5m'^2-m_V^2}{m^2-m'^2-
  2m_V^2}\frac{f^2_2}{f^2_1}
  \right\} \nonumber \\
  \end{equation}
  
  The prediction for $\alpha$ is independent of the value of the
  overlap integral and provide a test of the predictions (14) and
  (15) with $\xi_V = \xi_A$ through the presence of $f_2/f_1$.
  The corrections due to form factors which scales as $1/m$ are dumped
  into ${\cal F}$ functions
  
  If the vector meson $V$ is replaced by a pseudoscalar meson $P$, then
  \begin{equation}
  \rho_V(s) =0=\rho_A(s)
  \end{equation}
  \begin{equation}
  \sigma_A(s) = F_P^2\delta(s-m_P^2)
  \end{equation}
  where
  \begin{equation}
  <0|J'_\mu|p> =F_P q_\mu
  \end{equation}
  Then Eqs.(9) and (11), on using the relations (5), give
  \begin{equation}
  \Gamma_P = \frac{G'^2}{2} F_P^2  \frac{|{\bf p'}|}{4\pi m^2}
  Q'(m_P^2) [2f^2_1(m_P^2)]{\cal F}^P_1(m_P^2)
  \end{equation}
  \begin{equation}
  \alpha_P = \frac{-2m|{\bf p'}| \left[ (m^2 - m'^2)
  \right]}{2Q'(m_P^2)}
  \frac{{\cal F}^P_2(m_P^2)}{{\cal F}^P_1(m_P^2)}
  \end{equation}
  where
  \begin{equation}
  {\cal F}^P_1 (m_P^2) = \left\{1 - \frac{m'}{m} \frac{m_P^2(m^2-m'^2+
  m_P^2)}{Q'(m_P^2)} +
  \frac{m_P^2}{m^2} \frac{m_P^2(m^2+m'^2-m_P^2)}{2Q'(m_P^2)}
  \frac{f^2_2}{f^2_1}\right\}
  \end{equation}
  \begin{equation}
  {\cal F}^P_2(m_P^2) = \left\{1 -
  \frac{2m'}{m} \frac{m_P^2}{(m^2-m'^2)}\frac{f_2}{f_1} +
  \frac{m_P^4}{m^2(m^2-m'^2)}  \frac{f^2_2}{f^2_1}\right\}
  \end{equation}
  
  We are now ready to consider the specific decays. We first consider
  $\Lambda_b \longrightarrow \Lambda \, J/\psi$, where
  the first part of the Hamiltonian (1) with $q=c$ and Fierz
  rearrangement give
  \begin{equation}
  G' = G_F V_{cb} V^*_{cs} ( C_2 + \xi C_1)
  \end{equation}
  \begin{equation}
  J'_\mu = \bar{c} \gamma_\mu ( 1 + \gamma_5) c
  \end{equation}
  The constant $F^2_{J/\psi}$ is determined from
  $\Gamma(J/\psi \longrightarrow e^+ e^- ) = ( 5.26 \pm 0.37) \, KeV$
  \cite{R9}:
  \begin{eqnarray}
  F^2_{J/\psi} &=& \frac{9}{4} \left( \frac{3}{4 \pi \alpha^2} \right)
  \Gamma(J/\psi \longrightarrow e^+ e^- ) ( m_{J/\psi})\nonumber \\
  &=& 1.637 \times 10^{-1} \,\,\, GeV^2
  \end{eqnarray}
  Using $G_F = 1.16639\times 10^{-5} GeV^{-2}$ and \cite{R9}  $|V_{cb}|
  =
  0.0393 \pm 0.0028$, $|V_{cs}|=1.01\pm 0.18$, we obtain from Eqs.(23)
  and (24)
  \begin{eqnarray}
  \Gamma &=& 8.21 \times 10^{-14} (C_2+\xi C_1)^2 f_1^2 (m^2_{J/\psi})
  {\cal F}^V_1 (m^2_{J/\psi}) \\
  \alpha&=& -0.21 \frac{{\cal F}^V_2 (m^2_{J/\psi})}{{\cal F}^V_1
  (m^2_{J/\psi})}
  \end{eqnarray}
  This gives the branching ratio
  \begin{equation}
  \label{eq:22}
  B \left( \Lambda_b \longrightarrow \Lambda {J/\psi} \right)
  = 1.47 \times 10^{-1}   \left( C_2 + \xi C_1 \right)^2
  f_1^2 (m^2_{J/\psi})
  {\cal F}^V_1 (m^2_{J/\psi})
  \end{equation}
  where we have used \cite{R9} $\Gamma_{\Lambda_b} = 0.847\times
  10^{10}
  s^{-1} = 5.59\times 10^{-13} GeV$. Using Table 1
   we finally obtain
  \begin{equation}
  \label{eq:25}
  B\left( \Lambda_b \longrightarrow \Lambda {J/\psi}\right)
  = 9.14 \times 10^{-3} \left( C_2 +\xi C_1 \right)^2
  \end{equation}
  \begin{equation}
  \alpha \left( \Lambda_b \longrightarrow \Lambda {J/\psi}\right)
  = - 0.18
  \end{equation}
  In Fig.1, we show the branching ratio $B
  ( \Lambda_b \longrightarrow \Lambda {J/\psi} )$ as a function of
  $\xi$.
  This decay mode is sensitive to $\xi$ and comparison with the
  experimental
  value [13] $( 3.7 \pm 2.4 ) \times 10^{-4}$ shows that
  $\xi $ is restricted to $ 0  \leq \xi \leq 0.125$ or
  $0.35 \leq \xi \leq 0.45$,  which lie within the range $0\leq \xi
  \leq 0.5$ suggested by the combined analysis of the present
  CLEO data on $B \longrightarrow h_1 h_2 $ decay \cite{R2}.
  We may remark that $f_2/f_1$ correction to the decay rate is about
  6\%
  while that to the asymmetry parameter $\alpha $ is about $14 \%$.
  
  Other decays of interest for which the first part of Hamiltonian
  (\ref{eq:01}) with $q=c$ is responsible are
  $ \Lambda_b \longrightarrow \Lambda_c^+D^-_s$ and
  $ \Lambda_b \longrightarrow \Lambda_c^+D^{*-}_s$. For these decays
  \begin{equation}
  \label{eq:26}
  G' = {G_F} V_{cb} V^*_{cs} \left( C_1 + \xi C_2 \right)
  \end{equation}
  and
  \begin{equation}
  J'_\mu = \bar{s} \gamma_\mu ( 1 + \gamma_5) c
  \end{equation}
   Then Eqs.(23) and (24) and (29) and (30) [on using the relations
  (\ref{eq:05})]
  give respectively
  \begin{eqnarray}
  \Gamma \left( \Lambda_b \longrightarrow \Lambda_c^+D^{*-}_s \right)
  &=& 2.12 \times 10^{-14}
   \left( C_1 + \xi C_2 \right)^2
  f^2_1(m^2_{D^*_s} ) {\cal F}^V_1 (m^2_{D^*_s}) \nonumber \\ && \\
  \alpha \left( \Lambda_b \longrightarrow \Lambda_c^+D^{*-}_s \right)
  &=& - 0.42 \frac{ {\cal F}^V_2 (m^2_{D^*_s})}{{\cal F}^V_1
  (m^2_{D^*_s})} \\
  \label{eq:28}
  \Gamma \left( \Lambda_b \longrightarrow \Lambda_c^+D^-_s \right)
  &=& 1.50 \times 10^{-14}
   \left( C_1 + \xi C_2 \right)^2
  f^2_1(m^2_{D_s} ) {\cal F}^P_1 (m^2_{D_s}) \nonumber \\
  \end{eqnarray}
  \begin{equation}
  \alpha \left( \Lambda_b \longrightarrow \Lambda_c D_s \right)
  = - 0.98  \frac{ {\cal F}^P_2 (m^2_{D_s})}{{\cal F}^P_1
  (m^2_{D_s})}
  \end{equation}
  Here we have used $F_{D_s} = F_{D_s^*}=232\, MeV [12]$
  (in the normalization $F_\pi = 131 MeV) $. Using Table 1, the above
  equations give
  
  \begin{eqnarray}
  \label{eq:31}
  B \left( \Lambda_b \longrightarrow \Lambda_c D^*_s \right) &=&
  2.61 \left(C_1 + \xi C_2 \right)^2 \times 10^{-2}  \\
  \alpha \left( \Lambda_b \longrightarrow \Lambda_c D^*_s \right)
  &=& - 0.40 \\
  B \left( \Lambda_b \longrightarrow \Lambda_c D_s \right) &=&
  1.79 \left(C_1 + \xi C_2 \right)^2 \times 10^{-2} \\
  \alpha \left( \Lambda_b \longrightarrow \Lambda_c D_s \right)
  &=& - 0.98
  \end{eqnarray}
  The above branching ratios
  are not sensitive to $\xi$:  $2.55\times 10^{-2} \leq B(D^*_s)
  \leq 3.26 \times 10^{-2}$ and $1.75 \times 10^{-2} \leq B(D_s)
  \leq 2.23\times 10^{-2} $ for $0.5 \geq \xi \geq 0$. The $f_2/f_1$
  corrections are negligible when the meson in the final state is
  $O^-$  while for $1^-$
  they are about $5\%$  for the decay rate
  and for the asymmetry parameter $\alpha$.
  Previously the above decays have been analyzed
  in the heavy quark effective theory (HQET) with the factorization
  approximation in the large $N_c$ limit
  either by parameterising the Isgure-Wise form factor
  $G_1(v\cdot v')$ [c.f. Eq.(\ref{eq:05}) with $f_1 = G_1 +
  (m_{\Lambda_c} /
   m_{\Lambda_b}) G_2, \,\, f_2=- G_2 / m_{\Lambda_b},$ where since
  $\Lambda_c, \,\, \Lambda_b$ form a multiplet, the absence of second
  class currents implies $G_2 = 0$] \cite{R3} or
  by evaluating it in the large $N_c$ limit \cite{R4}.
  In contrast we have used  quark
  model to fix the baryonic form factors and Eqs.(15) and (16).  The
  comparison of our predicted results with
  the previous results mentioned above is presented in Table 2.
  \begin{table}
  \caption{Predictions for the branching ratios (BR) in $\%$ for
  $\Lambda_b \longrightarrow \Lambda^+_c D^{* -}_s $ and
  $\Lambda_b \longrightarrow \Lambda^+_c D_s $ in the large
  $N_c$ limit $(\xi=0)$.}
  
  \centerline{ \begin{tabular}{cccccccc}
  \hline \hline
  Decay processes &&& Present BR calculation &&& BR & BR \\
  &&&                 $(\xi =0)$             &&& Ref.\cite{R3} &
  Ref.\cite{R4}  \\ \hline
  $\Lambda_b \longrightarrow \Lambda^+_c D^{* -}_s $ &&& 3.26 &&&
  $1.73^{+0.20}_{-0.30}$ & 1.77 \\
  $\Lambda_b \longrightarrow \Lambda^+_c D^-_s $ &&& 2.23 &&&
  $2.30^{+0.30}_{-0.40}$ & 1.156 \\ \hline \hline
  \end{tabular} }
  \end{table}
  
  Finally we consider the decays
  $\Lambda_b \longrightarrow \Lambda \bar{D}^o$ and
  $\Lambda_b \longrightarrow p D_s $; the interest here
  is that the
  ratio of their decay widths
  with
  $\Lambda_b \longrightarrow \Lambda  \, J/\psi $ and
  $\Lambda_b \longrightarrow \Lambda_c D_s$ respectively
  can fix the  CKM-Wolfenstein
  parameter $(\rho^2 + \eta^2)$ or $|V_{ub}/V_{cb}|$, independent of
  $\xi$, where $\eta$
  indirectly determines CP-violation. For these   decays the
  second  part of the Hamiltonian(\ref{eq:01}) with $q=c$ [and Fierz
  rearrangement for the former] give
  \begin{eqnarray}
  \label{eq:32}
  \Gamma  \left( \Lambda_b \longrightarrow \Lambda \bar{D}^o \right)
  &=&  \left[ \frac{G_F}{\sqrt{2}} V_{ub} V^*_{cs} \left( C_2 +
  \xi C_1 \right)  \right]^2 \nonumber \\
  && \times \frac{2 |{\bf p'}| }{4\pi m^2_{\Lambda_b}} F_D^2
  \left[ f^{\Lambda D}_1
  \left(m^2_{D} \right) \right]^2  {\cal F}^P_1 \left(m^2_{D} \right)
  Q' \left(m^2_{D} \right) \nonumber \\
  && \\
  \Gamma  \left( \Lambda_b \longrightarrow p D_s \right)
  &=&  \left[ \frac{G_F}{\sqrt{2}} V_{ub} V^*_{cs} \left( C_1 +
  \xi C_2 \right)  \right]^2 \nonumber \\
  && \times \frac{2 |{\bf p'}| }{4\pi m^2_{\Lambda_b}} F_{D_s}^2
  \left[ f^{pD_s}_1
  \left(m^2_{D_s} \right) \right]^2 {\cal F}^P_1 \left(m^2_{D_s} \right)
  Q' \left(m^2_{D_s} \right) \nonumber \\
  \end{eqnarray}
  Using Table 1, $F_D = 200 \, MeV$ and taking into consideration
  differences in phase space
  factors $p', \, Q$ and $Q'$ we obtain
  \begin{equation}
  \frac{\Gamma \left( \Lambda_b \longrightarrow \Lambda \bar{D}^o
  \right)}{\Gamma \left( \Lambda_b \longrightarrow \Lambda J/\psi
   \right)} = 5.88 \times 10^{-2}
  \left| \frac{V_{ub} }{V_{cb} } \right|^2
  = 2.8 \times 10^{-3}  ( \rho^2 + \eta^2) \nonumber \\
  \end{equation}
  \begin{equation}
  \frac{\Gamma \left( \Lambda_b \longrightarrow p \, D_s
  \right)}{\Gamma \left( \Lambda_b \longrightarrow \Lambda_c \, D_s
   \right)} = 2 \times 10^{-2}
  \left| \frac{V_{ub} }{V_{cb} } \right|^2
  = 9.7 \times 10^{-4}  ( \rho^2 + \eta^2) \nonumber \\
  \end{equation}
  
  \section{CONCLUSIONS}
  We have analyzed some two body non-leptonic $\Lambda_b$
  decays in the factorization approximation, treating $\xi = 1/N_c$
  (which is supposed to compensate for the neglect of color octet-octet
  contribution in evaluating the hadronic matrix elements) as
  a free parameter.
  In addition we have used  quark model to fix baryonic form factors
  at the desired value of
  $s = - q^2$  without making no recoil approximation.
  The form factors obtained
  are consistent with the predictions of heavy quark symmetry and
  explicitly display $1/m_b$ corrections.
  The experimental branching ratio
  for $\Lambda_b \longrightarrow \Lambda {J/\psi}$  restricts $\xi$
  and can be understood for either $0 < \xi < 0.125$ or $0.3 <\xi<0.45$.
  Our predictions for the branching ratios
  $\Lambda_b \longrightarrow \Lambda_c D_s({D^*_s})$
  are larger  than the previous estimates. Future experimental data
  from colliders are expected
  to verify and distinguish the various results. Finally the parameter
  $|V_{ub}/V_{cb}|$ or $(\rho^2 + \eta^2)$
  can be determined independently of the parameter $\xi$ from the ratio
  of decay widths of
  $\Lambda_b \longrightarrow \Lambda \bar{D}$ and
  $\Lambda_b \longrightarrow \Lambda \, J/\psi$ or that of
  $\Lambda_b \longrightarrow p \, D_s$ and
  $\Lambda_b \longrightarrow \Lambda_c \, D_s$,
  although the branching ratios expected for these
  decays may be hard to measure.
  
      We want to emphasize that our derivation of Eqs. (16) and (17)
  does not
  depend on the details of quark model. The basic assumption is that in
  the
  heavy quark limit, the velocity of heavy quark can be neglected. The
  details of the quark model enter in the derivation of the overlap
  integral $I$.
  
      It may be noted from the structure of Eqs. (9) and (10), that the
  contribution of the form factors $f_{V}, \, h_{V},\, h_{A}$ and
  $f_{A}$ are
  proportional to $\frac{s}{m^{2}}$ [same is true for the term
  containing
   $(g^{2}_{V}-g^{2}_{A})$]. Hence when $\frac{s}{m^{2}}\, \ll \, 1$,
  their
   contribution can be neglected and in this case asymmetry parameter
  $\alpha$
   is given by
  \[ \alpha \, \simeq \, -\frac{2g_{V}g_{A}}{g^{2}_{V}+g^{2}_{A}}.\]
  
  \section*{ACKNOWLEDGMENTS}
  One of the authors (R) would like to acknowledge the support of King
  Fahd University of Petroleum and Minerals for this work and
  hospitality
  of the Theory Group at SLAC (during the summer of 1997) where a part
  of this work was done.
  
  \appendix
  \section{APPENDIX}
  We outline the derivation of relations (16) and (17). We first reduce
  the matrix elements in Eq.(4) from four component Dirac-spinors to
  Pauli spinors. Thus in the rest frame of $B_b$
  \[
  <B(p')|J_o|B_b(p)>
  \]
  \[= \sqrt{\frac{E' + m'}{2 E'}}
  \left\{ \left[ g_V(s) - q_o h_V(s) - \frac{{\bf q}^2}{E' + m'} f_V(s)
  \right] \right.
  \]
  \[
  + \left. \left[ h_A(s) + \frac{1}{E' + m'}
  \left( g_A (s) - q_o f_A(s) \right) \right] {\bf \sigma}
  \cdot {\bf q} \right\} \,\,\,\,\,\,\,\,\,\,\, (A-1)
  \]
  \[
  <B(p')|{\bf J} |B_b(p)>
  \]
  \[
  = \sqrt{\frac{E' + m'}{2 E'}}
  \left\{ \left[ - g_A(s) + \left(q_o  + \frac{{\bf q}^2}{E' + m'}
  \right) h_A(s) \right] {\bf \sigma} \right.
  \]
  \[
   - \left[ \left( 1 + \frac{q_o}{E' + m'} \right)f_V(s) +
  \frac{1}{E' + m'} g_V(s) \right] i {\bf \sigma } \times {\bf q}
  \]
  \[
   - \left[ h_V(s) + \frac{1}{E' + m'} \left( g_V(s) + q_o f_V(s)
  \right)
  \right] {\bf q}
  \]
  \[
   - \left. \frac{1}{E' + m'}  \left[ h_A(s) + f_A(s)
  \right] {\bf q} {\bf \sigma } \cdot {\bf q} \right\}
  \hspace{1in} \,\,\,\,\,\,\,\,\,\,\,\,(A-2)
  \]
  where $E'(s) = p'_o(s) $,  ${\bf q} = - {\bf p'}$,
  $q_o = \sqrt{ {\bf |q|}^2 + s}$. It may be noted that no approximation
  has been made so far. On the other hand, the Pauli reduction of the
  quark level current
  \[
  j_\mu = i {\bar{q}} \gamma_\mu \left( 1 + \gamma_5 \right) Q
  \hspace{2.6in}\,\,(A-3)
  \]
  is given by [ with $p_Q = p_3$, $p_q = p'_3$]:
  \[
  j_o = \frac{1}{2 \left[ E_3 E'_3 (E_3 + m_3) (E'_3 +m'_3)
  \right]^{1/2}}
  \]
  \[
   \times \left\{  (E'_3 + m'_3) (E_3 +m_3)  +
  {\bf p'_3} \cdot {\bf p_3} + i
  {\bf \sigma } \cdot ({\bf p'_3} \times {\bf p_3}  )
  \right.
  \]
  \[
   - \left. (E'_3 + m'_3){\bf \sigma} \cdot {\bf p_3}
  - (E_3 + m_3) {\bf \sigma} \cdot {\bf p'_3} \right\}
  \hspace{.5in} \,\,\,\,\,\,\,\,\, (A-4)
  \]
  \[
  {\bf j} = \frac{1}{2 \left[ E_3 E'_3 (E_3 + m_3) (E'_3 +m'_3)
  \right]^{1/2}}
  \]
  \[
   \times \left\{  \left[ - (E'_3 + m'_3) (E_3 +m_3)  +
  {\bf p'_3} \cdot {\bf p_3} \right] {\bf \sigma} + i
   ({\bf p'_3} \times {\bf p_3}  ) \right.
  \]
  \[
    - ({\bf \sigma} \cdot {\bf p'_3}) {\bf p_3}
     -   {\bf \sigma} \cdot {\bf p_3} {\bf p'_3}
  \]
  \[ +  (E'_3 + m'_3)\left( {\bf p_3} - i {\bf \sigma}
          \times  {\bf p_3} \right)
  \]
  \[
   +  \left. (E_3 + m_3)\left( {\bf p'_3} + i {\bf \sigma}
          \times  {\bf p'_3} \right) \right\}
  \hspace{2in}\,\,\,\,\,\,\,\,\, (A-5)
  \]
  We now treat the quark $Q$ extremely non-relativistically
  and thus put $|{\bf p}_3| \simeq 0$. Then
  \[
  j_o =
  \frac{1}{\sqrt{2  E'_3 (E'_3 +m'_3) }}
   \left\{  (E'_3 + m'_3)  - {\bf \sigma } \cdot {\bf p'_3}
  \right\}  \hspace{.7in} \,\,\,\,\,\, (A-6)
  \]
  \[
  {\bf j} =
  \frac{1}{\sqrt{2  E'_3 (E'_3 +m'_3) }}
   \left\{  - (E'_3 + m'_3) {\bf \sigma } + {\bf p'_3 } +
   i {\bf \sigma } \cdot {\bf p'_3}
  \right\}  \,\,\,\,\, (A-7)
  \]
  where $$ E'_3 = \sqrt{{\bf p'}^2_3 + m'^2_3} =
  \sqrt{({\bf p}_3 - {\bf q})^2 + m'^2_3} \simeq
  \sqrt{{\bf q}^2 + m'^2_3} $$
  
  Suppose that the initial baryon $B$ contains a heavy quark $Q$ ($b$
  in our case) and two light quarks $q_1$ and $q_2$ which behave as
  spectators. The final baryon $B'$ is composed of the quark $q$ [s, c,
  or u
  quark] and the same spectators as in $B$. For the initial baryon
  composed of quarks $Q( \equiv q_3)$, $q_1$, $q_2$, we introduce
  relative coordinates and momenta as
  \[
  {\bf r}_{12} = {\bf r}_1 - {\bf r}_2, \,\,
  \frac{{\bf p}_{12}}{m_{12}} = \frac{ {\bf p}_1 }{m_1} -
  \frac{ {\bf p}_2 }{m_2}, \,\,
  m_{12} = \frac{m_1 m_2}{m_1 + m_2}
  \]
  \[
  {\bf R}_{12} = \frac{ m_1 {\bf r}_{1} + m_2 {\bf r}_{2}}{m_{12}},
  \,\, {\bf r}_{12,3} = {\bf r}_{12} - {\bf r}_3,
  \]
  \[
  {\bf P}_{12} = {\bf p}_1 + {\bf p}_2, \,\,
  \frac{{\bf k} }{\mu} = \frac{ {\bf P}_{12} }{m_1 + m_2} -
  \frac{ {\bf p}_3 }{m_3}, \,\,
  \mu = \frac{m_3 (m_1 + m_2)}{\tilde{m}}
  \]
  \[
  \tilde{m} = m_1+m_2+m_3, \,\, {\bf k} = \frac{m_3}{\tilde{m}} {\bf
  P}_{12} -
  \frac{m_1 + m_2}{\tilde{m}} {\bf p}_3
  \hspace{0.2in} \,\,\,\,\,\, (A-8)
  \]
  For the initial baryons, its rest frame is its center of mass frame
  so that
  ${\bf p}_1 + {\bf p}_2 + {\bf p}_3 = 0$ which implies
  $ {\bf P}_{12} = - {\bf p}_3 = {\bf k}$ and then
  \[
  {\bf p}_1 = {\bf p}_{12} + \frac{m_1}{m_1 +m_2} {\bf k}
  \]
  \[
  {\bf p}_2 = - {\bf p}_{12} + \frac{m_1}{m_1 +m_2} {\bf k}
  \hspace{2.4in} (A-9)
  \]
  Denoting the relative momenta of quarks in the baryon $B'$ by primes
  and
  noting that $p'_1 = p_1$, $p'_2 =p_2$ so that ${\bf p}'_{12} = {\bf
  p}_{12}$,
  ${\bf P}'_{12} = {\bf P}_{12}$, giving ${\bf p'}_3 = - {\bf P}_{12}
  + {\bf p'} = - {\bf k} - {\bf q} $ and
  \[
  {\bf k'} = {\bf k} - \frac{m_1 +m_2}{\tilde{m}'} {\bf p'}
  \hspace{2.5in} (A-10)
  \]
  Calling $\psi_s$ the spatial wave function in momentum space and
  noting that when ${\bf p'}_3$ in Eqs.(A-6) and (A-7) is replaced by
  $- {\bf k} - {\bf q}$, the linear terms in ${\bf k}$
  do not contribute in the spatial integral and as such the right sides
  of Eqs.(A-6) and (A-7) are independent of integration variables
  ${\bf k}$, ${\bf p}_{12}$ and ${\bf k'}$. The comparison of
  hadronic matrix elements in Eqs.(A-1) and (A-2) with those of Eqs.(A-
  6) and (A-7) give the relations (16) and (17). The use of delta function
  $\delta ( {\bf p}_1 - {\bf p'}_1), \,\,
  \delta ( {\bf p}_2 - {\bf p'}_2), \,\,
  \delta ( {\bf p}_1 + {\bf p}_2 + {\bf p}_3) $ and
  $ ({\bf p'}_1 + {\bf p'}_2 + {\bf p'}_3 - {\bf p'})$
  reduce the spatial integral to the form given in Eq.(20).
  
  \newpage
  {\Huge Figure Captions}
  
  Figure 1. Branching ratio for $\Lambda_b \longrightarrow \Lambda
  {J/\psi}$
  as a function of $\xi$. The dotted and solid lines show the CDF
  measurement.
  
  \end{document}